\documentclass[aps,prl,twocolumn,superscriptaddress,groupedaddress]{revtex4}  

\usepackage{graphicx}  
\usepackage{dcolumn}   
\usepackage{bm}        
\usepackage{amssymb}   
\usepackage{epstopdf}
\usepackage{subfig}
\usepackage{amsmath}

\begin{document}

\widetext
\leftline{Version 11 as of \today}
\leftline{Primary authors: MIT, NTU, INFN Bari}
\leftline{Published in PRL}

\newcommand{\pt}{$p_{\text{T}}$}

\title{Measurements of Two-Particle Correlations in $e^+e^-$ Collisions at 91 GeV with ALEPH Archived Data}

\author{Anthony Badea}
\affiliation{Massachusetts Institute of Technology, Cambridge, Massachusetts, USA}%

\author{Austin Baty}
\affiliation{Massachusetts Institute of Technology, Cambridge, Massachusetts, USA}%

\author{Paoti Chang}
\affiliation{National Taiwan University, Taipei, Taiwan}%

\author{Gian Michele Innocenti}
\affiliation{Massachusetts Institute of Technology, Cambridge, Massachusetts, USA}%

\author{Marcello Maggi}
\affiliation{INFN Sezione di Bari, Bari, Italy}%

\author{Christopher McGinn}
\affiliation{Massachusetts Institute of Technology, Cambridge, Massachusetts, USA}%

\author{Michael Peters}
\affiliation{Massachusetts Institute of Technology, Cambridge, Massachusetts, USA}%

\author{Tzu-An Sheng}
\affiliation{National Taiwan University, Taipei, Taiwan}%

\author{Jesse Thaler}
\affiliation{Massachusetts Institute of Technology, Cambridge, Massachusetts, USA}%

\author{Yen-Jie Lee}
\email{yenjie@mit.edu}
\affiliation{Massachusetts Institute of Technology, Cambridge, Massachusetts, USA}%

\date{\today}

\begin{abstract}
Measurements of two-particle angular correlations of charged particles emitted in hadronic $Z$ decays are presented. The archived $e^+e^-$ annihilation data at a center-of-mass energy of 91 GeV were collected with the ALEPH detector at LEP between 1992 and 1995. The correlation functions are measured over a broad range of pseudorapidity and full azimuth as a function of charged particle multiplicity. No significant long-range correlation is observed in either the lab coordinate analysis or the thrust coordinate analysis, where the latter is sensitive to a medium expanding transverse to the color string between the outgoing $q\bar{q}$ pair from $Z$ boson decays. The associated yield distributions in both analyses are in better agreement with the prediction from the {\sc pythia} v6.1 event generator than from {\sc herwig} v7.1.5. They provide new insights to showering and hadronization modeling. These results serve as an important reference to the observed long-range correlation in proton-proton, proton-nucleus, and nucleus-nucleus collisions.
\end{abstract}

\maketitle

Measurements of two-particle angular correlation functions in high multiplicity proton-proton (pp), proton-nucleus (pA), and nucleus-nucleus (AA) collisions have revealed a ridgelike structure for particle pairs having large differences in pseudorapidity ($\Delta\eta$, where $\eta=-\ln{\tan{(\theta/2)}}$ and the polar angle $\theta$ is defined relative to the counterclockwise beam), but small differences in azimuthal angle ($\Delta\phi$)~\cite{Khachatryan:2010gv,Aad:2015gqa,CMS:2012qk,Abelev:2012ola,Aad:2012gla,Adare:2013piz,Chatrchyan:2012wg,Aaij:2015qcq}. In AA collisions, this long-range correlation is interpreted as a consequence of hydrodynamical expansion of the quark-gluon plasma with initial state fluctuations~\cite{Ollitrault:1992bk,Alver:2010gr}. However, the physical origin of the ridge signal in pp and pA collisions is not yet understood (see Refs.~\cite{Dusling:2015gta,Nagle:2018nvi} for recent reviews). Because of the complexity of hadron structure, possible initial state parton correlations could complicate the interpretation of pp and pA measurements. A large variety of theoretical models have been proposed to describe these high particle density systems, with underlying mechanisms ranging from initial state correlations~\cite{Dusling:2013qoz} to final-state interactions~\cite{He:2015hfa} and hydrodynamic effects~\cite{Bozek:2011if}. 

Unlike hadron-hadron collisions, electron-positron ($e^+e^-$) annihilations do not have beam remnants, gluonic initial state radiations, or the complications of parton distribution functions. Therefore, $e^+e^-$ collisions provide a cleaner environment than the more complex hadron systems previously considered. Since electrons and positrons are pointlike objects, no initial state correlation effects such as those from the possible formation of a color-glass condensate in hadrons contribute to the final state particle correlation functions. Furthermore, the initial momenta of the two quarks originating from $Z$ boson decays are fixed. The measurement of events with many final-state particles originating from the two-quark system could offer significant insights into the origin of the ridgelike signal~\cite{Nagle:2017sjv}.

This study uses archived data collected by the ALEPH detector at LEP~\cite{Decamp:1990jra} between 1992 and 1995. To analyze these data, an MIT Open Data format was created~\cite{Tripathee:2017ybi}. Hadronic events are selected by requiring the sphericity axis to have a polar angle in the laboratory reference frame ($\theta_{\text{lab}}$) between $7\pi/36$ and $29\pi/36$ to ensure that the event is well contained within the detector. At least five tracks having a minimum energy of 15 GeV are also required to suppress electromagnetic interactions~\cite{Barate:1996fi}. The residual contamination from processes such as $e^+e^-\rightarrow\tau^+\tau^-$ is expected to be less than 0.26\% for these event selections~\cite{Barate:1996fi}. Approximately 2.51 (2.44) million $e^+e^-$ collisions resulting in the decay of a $Z$ boson to quarks are analyzed (selected).

  High-quality tracks from particles are selected using requirements identical to those in previous ALEPH analyses~\cite{Barate:1996fi} and are also required to have a transverse momentum with respect to the beam axis ($p_{\rm T}^{\rm lab}$) above 0.2 GeV/c and $|\cos{\theta_{\text{lab}}}|<0.94$ in the lab frame. Secondary charged particles from neutral particle decays are suppressed by $V^0$ reconstruction in the energy flow algorithm~\cite{Barate:1996fi}. Archived $\textsc{pythia}$ 6.1~\cite{Sjostrand:2000wi} Monte Carlo (MC) simulation samples are used to derive efficiency correction factors for charged particles, and to correct detector effects and the contributions from the residual secondary particles which alter the correlation functions.  Event thrust distributions~\cite{Farhi:1977sg} published by the ALEPH Collaboration using a similar dataset~\cite{Heister:2003aj} were successfully reproduced within uncertainties, affirming that the archived data is analyzed properly. 
 
 The analysis is performed with a procedure similar to previous studies of two-particle correlation functions~\cite{CMS:2012qk}. For each event, the efficiency corrected differential yield of the number of charged-particle pairs ($\frac{\rm d^2 N^{\rm same}}{\rm d\Delta\eta \rm d\Delta\phi}$) is calculated.  Here the superscript ``same'' indicates that both particles in the pair come from the same event. This differential yield is scaled by the corrected number of charged particle tracks in the event ($\rm N_{\text{trk}}^{corr}$) averaged over all events of interest.  This forms the per-charged-particle yield of particle pairs:
 \begin{align}
    S(\Delta\eta,\Delta\phi) &= \frac{1}{\rm N_{\rm trk}^{corr}}\frac{\rm d^2 N^{\rm same}}{\rm d\Delta\eta \rm d\Delta\phi}.
\end{align}

\begin{figure}[t]%
    \centering
    \includegraphics[width=0.48\textwidth]{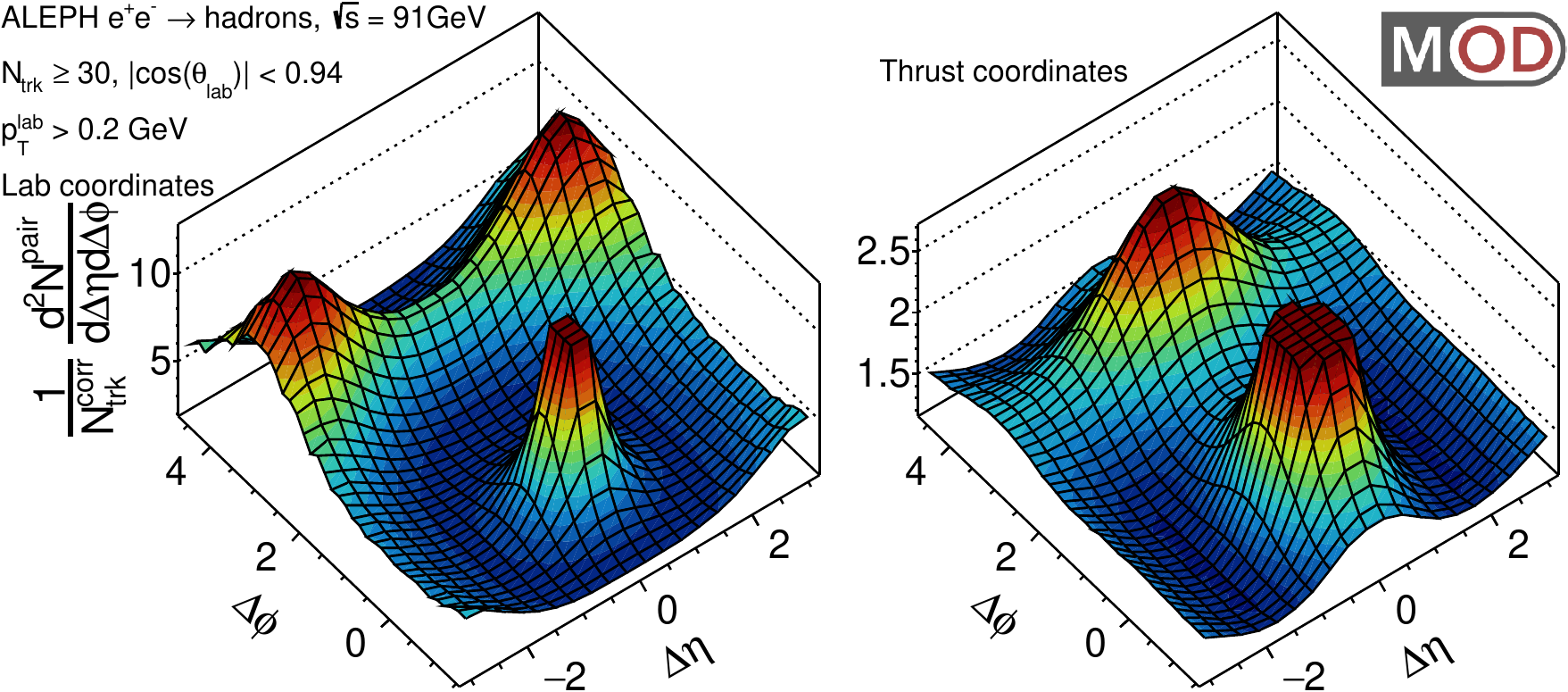}%
    \caption{Two-particle correlation functions for events with the number of charged particle tracks in the event $N_{\mathrm{trk}}\geq$ 30 in the lab coordinates (left) and thrust coordinates (right) analyses. The sharp near-side peaks arise from jet correlations and have been truncated to better illustrate the structure outside that region.}%
    \label{fig:2PC_BeamThrust}%
\end{figure}

A mixed-event background correlation function pairing the charged particles in one paired event with associated charged particles in 12 random events (5 in MC simulation studies) having the same multiplicity is also calculated:
 \begin{align}
   B(\Delta\eta,\Delta\phi) &= \frac{1}{\rm N_{trk}^{corr}}\frac{\rm d^2 N^{\rm mix}}{\rm d\Delta\eta \rm d\Delta\phi},
\end{align}
where $N^{\rm mix}$ denotes the efficiency corrected number of pairs taken from the mixed event. This mixed-event background correlation function, when divided by $B(0,0)$, represents the pair acceptance of the detector when particles in the pair are uncorrelated. Experimentally, $B(0,0)$ is calculated by using pairs with $|\Delta\eta|<0.32$ and $|\Delta\phi|<\pi/20$. Thus, the acceptance-corrected differential yield of particle pairs is given by
 \begin{align}
    \frac{1}{\rm N_{\rm trk}^{corr}}\frac{\rm d^2N^{pair}}{d\Delta\eta  \rm d\Delta\phi} &= B(0,0) \times \frac{S(\Delta\eta, \Delta\phi)}{B(\Delta\eta, \Delta\phi)}.
\end{align} \par
To study the event multiplicity dependence of the correlation function, the analysis is performed with events in 5 multiplicity intervals classified by the number of reconstructed charged particle tracks ($\rm N_\text{trk}$) with $p_{\rm T}^{\rm lab}>0.2$ GeV/c. The multiplicity ranges used, the corresponding fraction of the total sample, and the average number of tracks for each multiplicity class before ($\langle {\rm N_{\text{trk}}}\rangle$) and after detection efficiency correction ($\langle {\rm N_{\text{trk}}^{\text{corr}}}\rangle$) are summarized in Table~\ref{fig:table2}.

\begin{table}[t]\centering
\begin{tabular}{ cccc}
 \hline
 $\rm N_{\text{trk}}$ range & Fraction of data (\%) & $\langle {\rm N_{\text{trk}}}\rangle$ & $\langle {\rm N_{\text{trk}}^{\text{corr}}}\rangle$\\
 \hline
 $\left[ \mathrm{5,10} \right)$&   3.1& 8.2& 8.9\\
 $\left[ \mathrm{10,20} \right)$&   59.2& 15.2& 15.8\\
 $\left[ \mathrm{20,30} \right)$&   34.6& 23.1& 23.4\\
 $\left[ \mathrm{30,\infty} \right)$&   3.1& 32.4& 32.6\\
 $\left[ \mathrm{35,\infty} \right)$&   0.5& 36.9& 37.2\\
 \hline
\end{tabular}
\caption{Fraction of the full event sample for each multiplicity class. The last two columns
show the observed and corrected multiplicities, respectively, of charged particles with $p_{\rm T}^{\rm lab} >$ 0.2 GeV/c and $|\cos{\theta_{\text{lab}}}|<0.94$.}
\label{fig:table2}
\end{table}

The analysis is first performed with lab coordinates, similar to previous analyses at hadron colliders. In a hydrodynamics picture, the lab coordinate analysis is sensitive to the QCD medium expanding transverse to the beam axis. However, this coordinate system, although identical to what was used in the studies of heavy ion collisions, may not be the most suitable for the analysis of $e^+e^-$ collisions. Instead, using a coordinate system with the $z$ axis defined by the outgoing $q\bar{q}$ from the $Z$ decay enables a search for signal associated with the QCD medium expanding transverse to this direction. Experimentally, the thrust axis~\cite{Farhi:1977sg} is closely related to the outgoing $q\bar{q}$ direction and is used to define the coordinate system for the thrust coordinate analysis. For the purposes of calculating the thrust direction, an extra particle corresponding to the unreconstructed momentum of the event is included.  This reduces the effects of detector inefficiencies on the final correlation function.  Then every track passing quality selections has its kinematic variables (\pt, $\eta$, $\phi$) recalculated using the thrust axis to replace the role of the beam axis. The variation of the thrust axis direction causes the ALEPH detector acceptance in the thrust coordinates to vary on an event-by-event basis. This is accounted for by recalculating the kinematics for particles in paired events with respect to the thrust axis in the signal event. The $\eta$ and $\phi$ distributions of the charged tracks in the paired events are then reweighted to match that of signal events. 

The systematic uncertainty of the result is evaluated following a procedure similar to previous ALEPH studies~\cite{Barate:1996fi}. The required number of hits a track leaves in the ALEPH time projection chamber was varied from 4 to 7. From this variation, the tracking uncertainty is estimated to be 0.7\% in the lab coordinate analysis and 0.3\% in the thrust coordinate analysis. The hadronic event selection was studied by changing the required charged energy in an event to be 10 instead of 15 GeV. This only affects the lowest multiplicity bin, where an uncertainty of 0.6\% (3.4\%) is quoted for the lab (thrust) coordinate analysis.  A small correlated uncertainty of 0\%--0.1\% (0.1\%--0.9\%) on the value of $B(0,0)$ in the lab (thrust) coordinate analysis arising from statistical fluctuations is also included as a component of the systematic uncertainties. An additional systematic of 0.2\%-10\% (0.1\%-0.5\%) in the lab (thrust) coordinate analysis is included to quantify the residual uncertainty in the reconstruction effect correction factor derived from the {\sc pythia 6.1} archived MC sample, which is mainly from the limited size of the archived MC sample. In general, the systematic uncertainties in thrust analysis are smaller than the beam axis analysis because the thrust correlation function before the combinatorial background subtraction described later is quite flat, and variations affecting the correlation shape are less pronounced.

The two-particle correlation functions for events with $\rm N_{\text{trk}}\geq30$ are shown in Fig.~\ref{fig:2PC_BeamThrust}. The left panel shows the correlation function using lab coordinates, while the right panel shows the result when using thrust coordinates. In both cases, the dominant feature is the jet peak near $(\Delta\eta,\Delta\phi)=(0,0)$ arising from particle pairs within the same jet. For the analysis using lab coordinates, the away-side structure at $\Delta\phi\sim\pi$ arises from pairs of particles contained in back-to-back jets. In the thrust coordinate analysis, this peaking structure is related to multijet topologies. For instance, the thrust axis points to the direction of the leading jet in a three-jet event and the correlation between the particles in the subleading and third jet can create a narrow peak at small $\Delta\eta$ and at $\Delta\phi\sim\pi$. Because many charged particles are approximately aligned with the thrust axis, i.e., at very large $\eta$ in the thrust coordinate, particle pairs in back-to-back jets frequently have a $\Delta\eta$ larger than the $\Delta\eta$ range examined here, and do not contribute the correlation function in the analyzed $\Delta\eta$ window. This reduces the absolute magnitude of the correlation function in the thrust coordinate analysis compared to that in the lab coordinate analysis. Unlike previous results from hadron collisions, no significant ``ridge" structure is found around $\Delta\phi =0$ in either the lab or the thrust coordinate analysis. 


\begin{figure}[tb]
\begin{center}
\includegraphics[width=.48\textwidth]{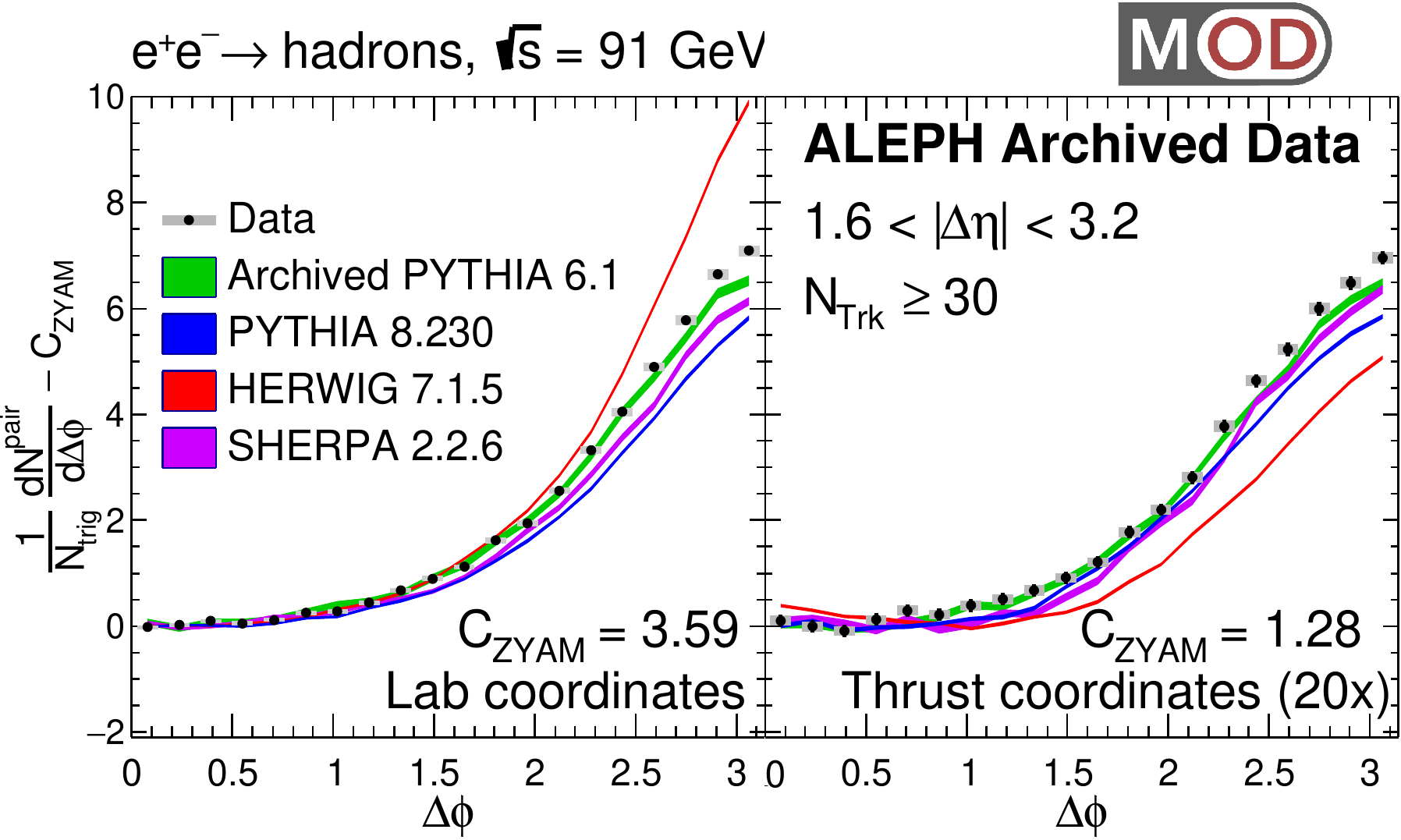}
\caption{Correlated yield obtained from the ZYAM procedure as a function of $|\Delta\phi |$ averaged over $1.6 < |\Delta\eta| < 3.2$ in lab (left) and thrust (right) coordinate analyses. Statistical
uncertainties are smaller than the marker size and the systematic uncertainties are shown as gray boxes. The subtracted ZYAM constant for the data is listed in each panel.  Unlike the data points, the thrust ZYAM constant has not been scaled by a factor of 20.}
\label{fig:dNdphi} 
\end{center}
\end{figure}

To investigate the long-range correlation in finer detail, one-dimensional distributions in $\Delta\phi$ are found by averaging two-particle correlation function over the region between $1.6 < |\Delta\eta| < 3.2$.  The size of any potential enhancement around $\Delta\phi = 0$ is calculated by fitting this distribution from $0 < \Delta\phi < \pi/2$ and then performing a zero yield at minimum (ZYAM) subtraction procedure using the fit minimum, $c_{\text{ZYAM}}$~\cite{Ajitanand:2005jj}. A constant plus a three term Fourier series was used as the nominal fit function, but a fourth degree polynomial fit and a third degree polynomial plus a $\cos{2\Delta\phi}$ term fit were also attempted.  Discrepancies resulting from these different choices of fit function were found to be small and are included in the systematic uncertainties of the total near-side yield calculation. The results after this subtraction and correction for reconstruction effects are shown for $N_{\rm trk} \geq 30$ in Fig.~\ref{fig:dNdphi}. Because of the relatively small associated yield, the results from thrust coordinates are scaled by a factor of 20 for visual clarity.
A peak structure is observed at $\Delta\phi = \pi$ in both lab and thrust coordinate analyses, but the spectra decrease to values consistent with zero at $\Delta\phi = 0$. 
To test the impact of the perturbative and nonperturbative aspects of the implementation in MC event generators, these results are compared to calculations from {\sc pythia} v6.1~\cite{Sjostrand:2000wi} (from archived MC), {\sc pythia} v8.230~\cite{Sjostrand:2014zea}, {\sc herwig} v7.1.5~\cite{Bellm:2015jjp,Reichelt:2017hts} and {\sc sherpa} v2.2.6~\cite{Gleisberg:2008ta}. Both {\sc pythia} versions use a Lund string hadronization model, whereas {\sc sherpa} and {\sc herwig} implement cluster hadronization. The predictions from the {\sc pythia} v6.1 model, which was tuned to describe the ALEPH data, give the best description of the data. Both {\sc pythia} v8.230 and {\sc sherpa} v2.2.6 slightly underpredict the magnitude of the peak at $\Delta\phi = \pi$. The data are incompatible with the prediction from {\sc herwig}. Unlike the results with high multiplicity selection, all four generators studied were able to reproduce the lab coordinate correlation function in the 10--20 multiplicity bin and are therefore expected to give a reasonable model of inclusive $e^{+}e^{-}$ collisions.



\begin{figure}[!t]
\begin{center}
\includegraphics[width=.48\textwidth]{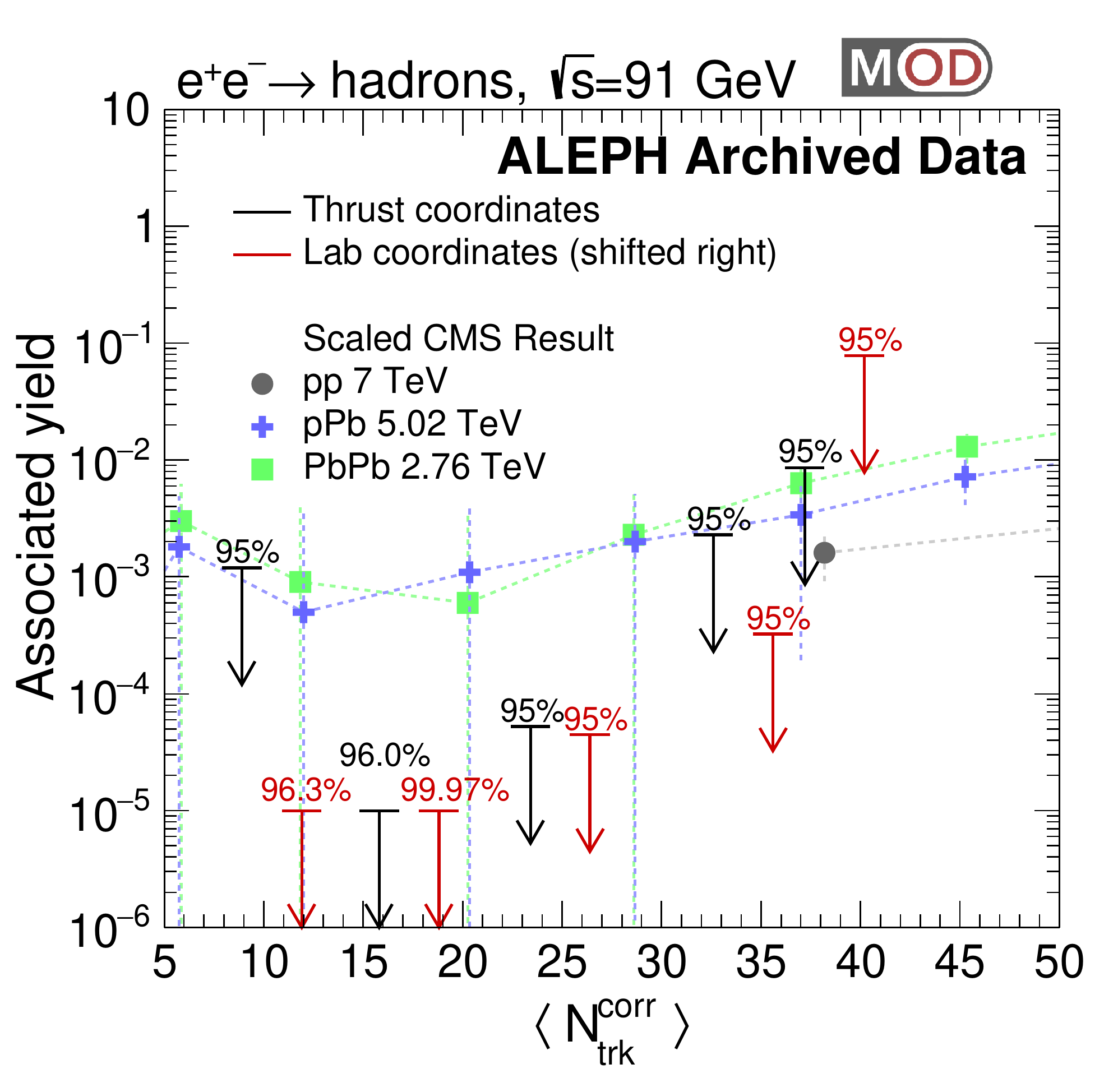}
\caption{Confidence limits on associated yield as a function of $\left \langle {\rm N}_{\mathrm{trk}}^{\mathrm{corr}} \right \rangle$. The results from lab (thrust) coordinates are shown as red (black) arrows. The lab data have been shifted right three units for clarity.  Scaled results (see text) from pp, pPb and PbPb collisions are shown as black circles, blue crosses and green boxes.}
\label{fig:confidenceLimitsBothAxis} 
\end{center}
\end{figure}

The total size of any excess yield of particle pairs around $\Delta\phi = 0$ is quantified by integrating the data from $\Delta\phi$ = 0 to the position in $\Delta\phi$ of the ZYAM fit's minimum. In general, no significant enhancement of particle pairs is observed in any of the multiplicity bins examined for either the lab or the thrust coordinate analysis. Therefore, a confidence limit (C.L.) on the near-side excess of particle pairs is calculated using a bootstrap procedure~\cite{efron1979}.  This method calculates the distribution of the associated yield after allowing the one-dimensional correlation function data points to vary according to their uncertainties. For each ${\rm N}_{\rm trk}$ bin, $2 \times 10^5$ variations were sampled in the bootstrap procedure. Most of these variations result in a correlation function that has a minimum at $\Delta \phi = 0$ and therefore zero associated yield.  If more than 5\% of the data variations have a yield above $1\times 10^{-5}$, a 95\% C.L. is quoted.  Otherwise, a C.L. corresponding to the fraction of data variations having a yield below $1\times 10^{-5}$ is reported.  This occurs in the low multiplicity selections, where the small uncertainties make it extremely unlikely that a bootstrap variation produces any nonzero associated yield.  The C.L.s are shown as a function of $\left \langle {\rm N}_{\mathrm{trk}}^{\mathrm{corr}} \right \rangle$ in Fig.~\ref{fig:confidenceLimitsBothAxis} by the red arrows for the lab coordinate analysis and black arrows for the thrust coordinate analysis.  In general, the constraining power of the data is driven mainly by statistical uncertainties, with multiplicity bins having more events also having lower C.L.s. The results are also compared to the associated yield measurements in pp, pPb and PbPb collsions reported by CMS ~\cite{Khachatryan:2010gv,CMS:2012qk,Khachatryan:2015lva}, where the $x$ axis of the CMS data was scaled by the pseudorapidity acceptance ratio between ALEPH and CMS (0.725) and corrected for the CMS minimum-bias tracking inefficiency in pp collisions (a factor of 1.15). The reported thrust C.L.s are compatible or lower than the central values of the associated yield reported by CMS, although the systematic uncertainties of the CMS measurements at low multiplicity are large. These C.L.s contrast measurements of a nonzero azimuthal anisotropy signal in lower multiplicity pp collisions~\cite{Aaboud:2016yar,Khachatryan:2016txc}.

\begin{figure}[!t]
\begin{center}
\includegraphics[width=.48\textwidth]{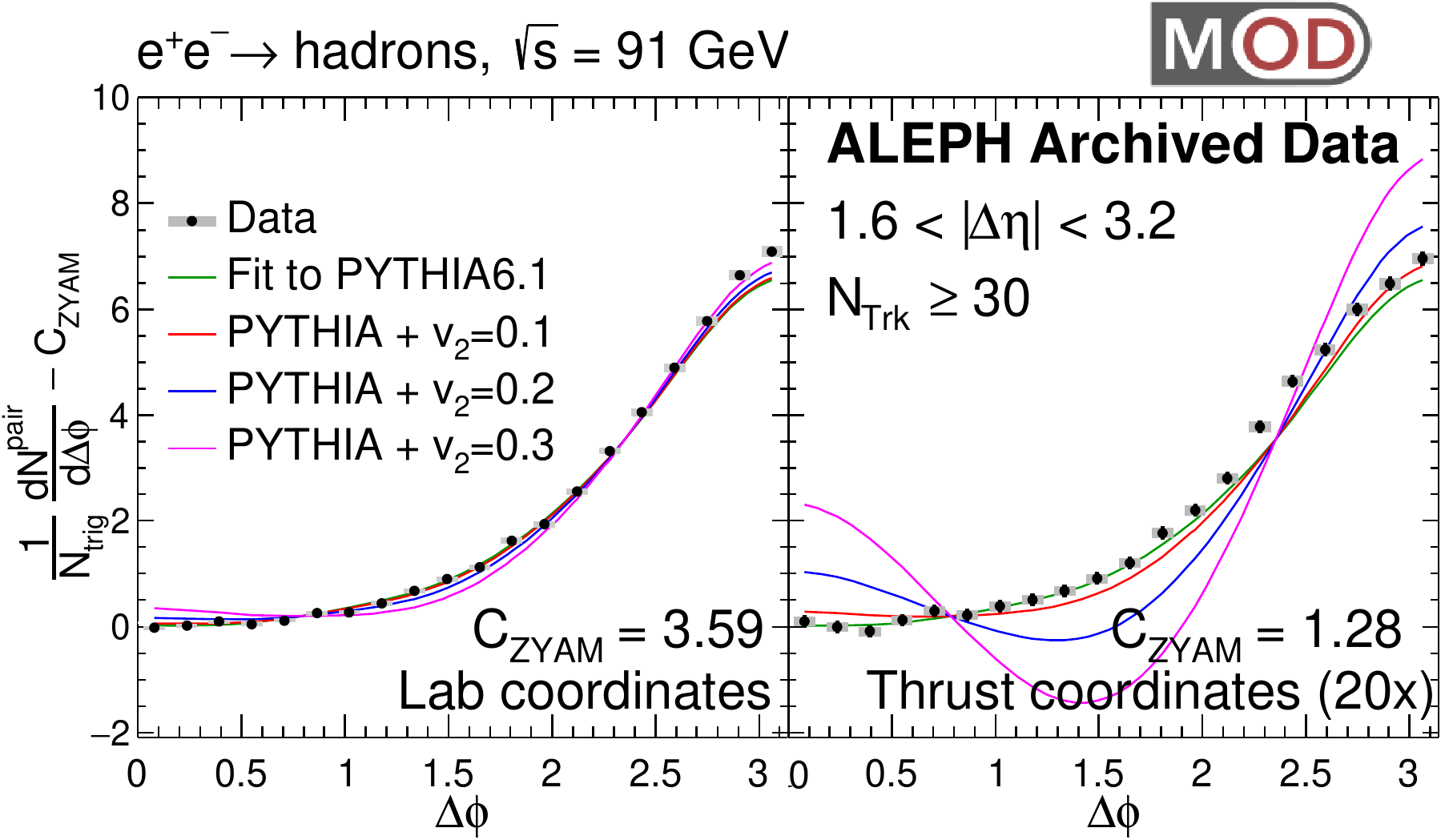}
\caption{ A comparison of the data to three simple \textsc{pythia}-based models that assume progressively larger values of $v_{2}$.}
\label{fig:v2Comparison} 
\end{center}
\end{figure}

In hadronic collision systems, the azimuthal anisotropy of charged particle production is typically quantified with the azimuthal anisotropy coefficients, $v_{n}$~\cite{Voloshin:1994mz,Poskanzer:1998yz,Alver:2010gr}.  In particular, the second order coefficient, $v_{2}$ is sensitive to the collective behavior and the level of thermalization of the system in relativistic heavy ion collisions~\cite{Ollitrault:1992bk,Ackermann:2000tr}. However, it is often difficult to make a direct quantitative connection between the size of any associated yields and the corresponding value of $v_{2}$ because most of the structure of the correlation functions comes from jetlike correlations.  These correlations are sometimes referred to as ``nonflow''~\cite{Adler:2002tq,Adare:2008ae,Aamodt:2010pa,Sirunyan:2017pan}.  To give an idea of the sensitivity of this analysis to nonzero values of $v_{2}$, a simple model was constructed using the archived \textsc{pythia} v6.1 as a baseline.  In this model, it is assumed that the portion of the one-dimensional correlation function that is subtracted by the ZYAM procedure could have an additional azimuthal modulation of 10, 20, or 30\%.  The new one-dimensional correlation function, after adding this additional $v_{2}$ component, is shown by the red, blue, and pink lines in Fig.~\ref{fig:v2Comparison}. Under this assumption that only the ZYAM-subtracted portion of the correlation could have an additional $v_{2}$ component, it appears that the measurement in the lab coordinates is not compatible with $v_{2}$ values of 0.2 or 0.3, but could perhaps still be compatible with $v_{2}=0.1$.  In the thrust axis, the spacing between the different $v_{2}$ assumptions is larger because the size of the ZYAM-subtracted part of the correlation is a much greater component of the total correlation function. Thus, the data have greater constraining power.  It should be emphasized that the conclusions mentioned here are strongly dependent on the assumption on the fraction of nonflow and flowlike contributions to the total correlation function.  It is possible that the nonflow contributions are even larger than what is assumed here, which would cause the analysis to be less sensitive to the precise value of $v_{2}$, but also decrease the importance of any flowlike effects in the total behavior of the system.  

In summary, the first measurement of two-particle angular correlations for charged particles emitted in $e^+e^-$ collisions at a center-of-mass energy of 91 GeV is reported using archived data collected with the ALEPH detector at LEP. The correlation functions are measured over a broad range of pseudorapidity and azimuthal angle of the charged particles. Those results using either lab coordinates or the event thrust coordinates are compared to predictions from the $\textsc{pythia}$, {\sc sherpa} and {\sc herwig} event generators. In contrast to the results from high multiplicity pp, pA and AA collisions, where long-range correlations with large pseudorapidity gap are observed, no significant enhancement of long-range correlations is observed in $e^+e^-$ collisions. The data are compared to generators that do not include additional final-state interactions of the outgoing partons. The results are better described by the {\sc pythia} and {\sc sherpa} generators than {\sc herwig}.  Those results provide new insights to the showering and hadronization modeling and serve as an important reference to the observed long-range correlation in high multiplicity pp, pA and AA collisions.

The authors would like to thank the ALEPH Collaboration for their support and foresight in archiving their data. We would like to thank the useful comments and suggestions from Roberto Tenchini, Guenther Dissertori, Wei Li, Maxime Guilbaud, Wit Busza, Yang-Ting Chien, Camelia Mironov, Fuqiang Wang, N\'estor Armesto, Jean-Yves Ollitrault, J\"urgen Schukraft and Jan Fiete Grosse-Oetringhaus. This work has been supported by the Department of Energy, Office of Science, under Grant No. DE-SC0013905 (to A. Baty, C.M., G.I. and M.P.), Grant No. DE-SC0012567 (to J.T.), the Department of Energy under Early Career Award no. DE-SC0013905 (to Y.L.), the MIT UROP office (to A. Badea), the Ministry of Education and the Ministry of Science and Technology of Taiwan (to P.C. and T.S.).

\bibliography{ridgepaperALEPH}

\begin{thebibliography}{38}
\expandafter\ifx\csname natexlab\endcsname\relax\def\natexlab#1{#1}\fi
\expandafter\ifx\csname bibnamefont\endcsname\relax
  \def\bibnamefont#1{#1}\fi
\expandafter\ifx\csname bibfnamefont\endcsname\relax
  \def\bibfnamefont#1{#1}\fi
\expandafter\ifx\csname citenamefont\endcsname\relax
  \def\citenamefont#1{#1}\fi
\expandafter\ifx\csname url\endcsname\relax
  \def\url#1{\texttt{#1}}\fi
\expandafter\ifx\csname urlprefix\endcsname\relax\def\urlprefix{URL }\fi
\providecommand{\bibinfo}[2]{#2}
\providecommand{\eprint}[2][]{\url{#2}}

\bibitem[{\citenamefont{Khachatryan et~al.}(2010)}]{Khachatryan:2010gv}
\bibinfo{author}{\bibfnamefont{V.}~\bibnamefont{Khachatryan}}
  \bibnamefont{et~al.} (\bibinfo{collaboration}{CMS}), \bibinfo{journal}{JHEP}
  \textbf{\bibinfo{volume}{09}}, \bibinfo{pages}{091} (\bibinfo{year}{2010}),
  \eprint{1009.4122}.

\bibitem[{\citenamefont{Aad et~al.}(2016)}]{Aad:2015gqa}
\bibinfo{author}{\bibfnamefont{G.}~\bibnamefont{Aad}} \bibnamefont{et~al.}
  (\bibinfo{collaboration}{ATLAS}), \bibinfo{journal}{Phys. Rev. Lett.}
  \textbf{\bibinfo{volume}{116}}, \bibinfo{pages}{172301}
  (\bibinfo{year}{2016}), \eprint{1509.04776}.

\bibitem[{\citenamefont{Chatrchyan et~al.}(2013)}]{CMS:2012qk}
\bibinfo{author}{\bibfnamefont{S.}~\bibnamefont{Chatrchyan}}
  \bibnamefont{et~al.} (\bibinfo{collaboration}{CMS}), \bibinfo{journal}{Phys.
  Lett.} \textbf{\bibinfo{volume}{B718}}, \bibinfo{pages}{795}
  (\bibinfo{year}{2013}), \eprint{1210.5482}.

\bibitem[{\citenamefont{Abelev et~al.}(2013)}]{Abelev:2012ola}
\bibinfo{author}{\bibfnamefont{B.}~\bibnamefont{Abelev}} \bibnamefont{et~al.}
  (\bibinfo{collaboration}{ALICE}), \bibinfo{journal}{Phys. Lett.}
  \textbf{\bibinfo{volume}{B719}}, \bibinfo{pages}{29} (\bibinfo{year}{2013}),
  \eprint{1212.2001}.

\bibitem[{\citenamefont{Aad et~al.}(2013)}]{Aad:2012gla}
\bibinfo{author}{\bibfnamefont{G.}~\bibnamefont{Aad}} \bibnamefont{et~al.}
  (\bibinfo{collaboration}{ATLAS}), \bibinfo{journal}{Phys. Rev. Lett.}
  \textbf{\bibinfo{volume}{110}}, \bibinfo{pages}{182302}
  (\bibinfo{year}{2013}), \eprint{1212.5198}.

\bibitem[{\citenamefont{Adare et~al.}(2013)}]{Adare:2013piz}
\bibinfo{author}{\bibfnamefont{A.}~\bibnamefont{Adare}} \bibnamefont{et~al.}
  (\bibinfo{collaboration}{PHENIX}), \bibinfo{journal}{Phys. Rev. Lett.}
  \textbf{\bibinfo{volume}{111}}, \bibinfo{pages}{212301}
  (\bibinfo{year}{2013}), \eprint{1303.1794}.

\bibitem[{\citenamefont{Chatrchyan et~al.}(2012)}]{Chatrchyan:2012wg}
\bibinfo{author}{\bibfnamefont{S.}~\bibnamefont{Chatrchyan}}
  \bibnamefont{et~al.} (\bibinfo{collaboration}{CMS}), \bibinfo{journal}{Eur.
  Phys. J.} \textbf{\bibinfo{volume}{C72}}, \bibinfo{pages}{2012}
  (\bibinfo{year}{2012}), \eprint{1201.3158}.

\bibitem[{\citenamefont{Aaij et~al.}(2016)}]{Aaij:2015qcq}
\bibinfo{author}{\bibfnamefont{R.}~\bibnamefont{Aaij}} \bibnamefont{et~al.}
  (\bibinfo{collaboration}{LHCb}), \bibinfo{journal}{Phys. Lett.}
  \textbf{\bibinfo{volume}{B762}}, \bibinfo{pages}{473} (\bibinfo{year}{2016}),
  \eprint{1512.00439}.

\bibitem[{\citenamefont{Ollitrault}(1992)}]{Ollitrault:1992bk}
\bibinfo{author}{\bibfnamefont{J.-Y.} \bibnamefont{Ollitrault}},
  \bibinfo{journal}{Phys. Rev.} \textbf{\bibinfo{volume}{D46}},
  \bibinfo{pages}{229} (\bibinfo{year}{1992}).

\bibitem[{\citenamefont{Alver and Roland}(2010)}]{Alver:2010gr}
\bibinfo{author}{\bibfnamefont{B.}~\bibnamefont{Alver}} \bibnamefont{and}
  \bibinfo{author}{\bibfnamefont{G.}~\bibnamefont{Roland}},
  \bibinfo{journal}{Phys. Rev.} \textbf{\bibinfo{volume}{C81}},
  \bibinfo{pages}{054905} (\bibinfo{year}{2010}), \bibinfo{note}{[Erratum:
  Phys. Rev.C82,039903(2010)]}, \eprint{1003.0194}.

\bibitem[{\citenamefont{Dusling et~al.}(2016)\citenamefont{Dusling, Li, and
  Schenke}}]{Dusling:2015gta}
\bibinfo{author}{\bibfnamefont{K.}~\bibnamefont{Dusling}},
  \bibinfo{author}{\bibfnamefont{W.}~\bibnamefont{Li}}, \bibnamefont{and}
  \bibinfo{author}{\bibfnamefont{B.}~\bibnamefont{Schenke}},
  \bibinfo{journal}{Int. J. Mod. Phys.} \textbf{\bibinfo{volume}{E25}},
  \bibinfo{pages}{1630002} (\bibinfo{year}{2016}), \eprint{1509.07939}.

\bibitem[{\citenamefont{Nagle and Zajc}(2018)}]{Nagle:2018nvi}
\bibinfo{author}{\bibfnamefont{J.~L.} \bibnamefont{Nagle}} \bibnamefont{and}
  \bibinfo{author}{\bibfnamefont{W.~A.} \bibnamefont{Zajc}},
  \bibinfo{journal}{Ann. Rev. Nucl. Part. Sci.} \textbf{\bibinfo{volume}{68}},
  \bibinfo{pages}{211} (\bibinfo{year}{2018}), \eprint{1801.03477}.

\bibitem[{\citenamefont{Dusling and Venugopalan}(2013)}]{Dusling:2013qoz}
\bibinfo{author}{\bibfnamefont{K.}~\bibnamefont{Dusling}} \bibnamefont{and}
  \bibinfo{author}{\bibfnamefont{R.}~\bibnamefont{Venugopalan}},
  \bibinfo{journal}{Phys. Rev.} \textbf{\bibinfo{volume}{D87}},
  \bibinfo{pages}{094034} (\bibinfo{year}{2013}), \eprint{1302.7018}.

\bibitem[{\citenamefont{He et~al.}(2016)\citenamefont{He, Edmonds, Lin, Liu,
  Molnar, and Wang}}]{He:2015hfa}
\bibinfo{author}{\bibfnamefont{L.}~\bibnamefont{He}},
  \bibinfo{author}{\bibfnamefont{T.}~\bibnamefont{Edmonds}},
  \bibinfo{author}{\bibfnamefont{Z.-W.} \bibnamefont{Lin}},
  \bibinfo{author}{\bibfnamefont{F.}~\bibnamefont{Liu}},
  \bibinfo{author}{\bibfnamefont{D.}~\bibnamefont{Molnar}}, \bibnamefont{and}
  \bibinfo{author}{\bibfnamefont{F.}~\bibnamefont{Wang}},
  \bibinfo{journal}{Phys. Lett.} \textbf{\bibinfo{volume}{B753}},
  \bibinfo{pages}{506} (\bibinfo{year}{2016}), \eprint{1502.05572}.

\bibitem[{\citenamefont{Bozek}(2012)}]{Bozek:2011if}
\bibinfo{author}{\bibfnamefont{P.}~\bibnamefont{Bozek}},
  \bibinfo{journal}{Phys. Rev.} \textbf{\bibinfo{volume}{C85}},
  \bibinfo{pages}{014911} (\bibinfo{year}{2012}), \eprint{1112.0915}.

\bibitem[{\citenamefont{Nagle et~al.}(2018)\citenamefont{Nagle, Belmont, Hill,
  Koop, Perepelitsa, Yin, Lin, and McGlinchey}}]{Nagle:2017sjv}
\bibinfo{author}{\bibfnamefont{J.~L.} \bibnamefont{Nagle}},
  \bibinfo{author}{\bibfnamefont{R.}~\bibnamefont{Belmont}},
  \bibinfo{author}{\bibfnamefont{K.}~\bibnamefont{Hill}},
  \bibinfo{author}{\bibfnamefont{J.~O.} \bibnamefont{Koop}},
  \bibinfo{author}{\bibfnamefont{D.~V.} \bibnamefont{Perepelitsa}},
  \bibinfo{author}{\bibfnamefont{P.}~\bibnamefont{Yin}},
  \bibinfo{author}{\bibfnamefont{Z.-W.} \bibnamefont{Lin}}, \bibnamefont{and}
  \bibinfo{author}{\bibfnamefont{D.}~\bibnamefont{McGlinchey}},
  \bibinfo{journal}{Phys. Rev. C} \textbf{\bibinfo{volume}{97}},
  \bibinfo{pages}{024909} (\bibinfo{year}{2018}),
  \urlprefix\url{https://link.aps.org/doi/10.1103/PhysRevC.97.024909}.

\bibitem[{\citenamefont{Decamp et~al.}(1990)}]{Decamp:1990jra}
\bibinfo{author}{\bibfnamefont{D.}~\bibnamefont{Decamp}} \bibnamefont{et~al.}
  (\bibinfo{collaboration}{ALEPH}), \bibinfo{journal}{Nucl. Instrum. Meth.}
  \textbf{\bibinfo{volume}{A294}}, \bibinfo{pages}{121} (\bibinfo{year}{1990}),
  \bibinfo{note}{[Erratum: Nucl. Instrum. Meth.A303,393(1991)]}.

\bibitem[{\citenamefont{Tripathee et~al.}(2017)\citenamefont{Tripathee, Xue,
  Larkoski, Marzani, and Thaler}}]{Tripathee:2017ybi}
\bibinfo{author}{\bibfnamefont{A.}~\bibnamefont{Tripathee}},
  \bibinfo{author}{\bibfnamefont{W.}~\bibnamefont{Xue}},
  \bibinfo{author}{\bibfnamefont{A.}~\bibnamefont{Larkoski}},
  \bibinfo{author}{\bibfnamefont{S.}~\bibnamefont{Marzani}}, \bibnamefont{and}
  \bibinfo{author}{\bibfnamefont{J.}~\bibnamefont{Thaler}},
  \bibinfo{journal}{Phys. Rev.} \textbf{\bibinfo{volume}{D96}},
  \bibinfo{pages}{074003} (\bibinfo{year}{2017}), \eprint{1704.05842}.

\bibitem[{\citenamefont{Barate et~al.}(1998)}]{Barate:1996fi}
\bibinfo{author}{\bibfnamefont{R.}~\bibnamefont{Barate}} \bibnamefont{et~al.}
  (\bibinfo{collaboration}{ALEPH}), \bibinfo{journal}{Phys. Rept.}
  \textbf{\bibinfo{volume}{294}}, \bibinfo{pages}{1} (\bibinfo{year}{1998}).

\bibitem[{\citenamefont{Sjostrand et~al.}(2001)\citenamefont{Sjostrand, Eden,
  Friberg, Lonnblad, Miu, Mrenna, and Norrbin}}]{Sjostrand:2000wi}
\bibinfo{author}{\bibfnamefont{T.}~\bibnamefont{Sjostrand}},
  \bibinfo{author}{\bibfnamefont{P.}~\bibnamefont{Eden}},
  \bibinfo{author}{\bibfnamefont{C.}~\bibnamefont{Friberg}},
  \bibinfo{author}{\bibfnamefont{L.}~\bibnamefont{Lonnblad}},
  \bibinfo{author}{\bibfnamefont{G.}~\bibnamefont{Miu}},
  \bibinfo{author}{\bibfnamefont{S.}~\bibnamefont{Mrenna}}, \bibnamefont{and}
  \bibinfo{author}{\bibfnamefont{E.}~\bibnamefont{Norrbin}},
  \bibinfo{journal}{Comput. Phys. Commun.} \textbf{\bibinfo{volume}{135}},
  \bibinfo{pages}{238} (\bibinfo{year}{2001}), \eprint{hep-ph/0010017}.

\bibitem[{\citenamefont{Farhi}(1977)}]{Farhi:1977sg}
\bibinfo{author}{\bibfnamefont{E.}~\bibnamefont{Farhi}},
  \bibinfo{journal}{Phys. Rev. Lett.} \textbf{\bibinfo{volume}{39}},
  \bibinfo{pages}{1587} (\bibinfo{year}{1977}).

\bibitem[{\citenamefont{Heister et~al.}(2004)}]{Heister:2003aj}
\bibinfo{author}{\bibfnamefont{A.}~\bibnamefont{Heister}} \bibnamefont{et~al.}
  (\bibinfo{collaboration}{ALEPH}), \bibinfo{journal}{Eur. Phys. J.}
  \textbf{\bibinfo{volume}{C35}}, \bibinfo{pages}{457} (\bibinfo{year}{2004}).

\bibitem[{\citenamefont{Ajitanand et~al.}(2005)\citenamefont{Ajitanand,
  Alexander, Chung, Holzmann, Issah, Lacey, Shevel, Taranenko, and
  Danielewicz}}]{Ajitanand:2005jj}
\bibinfo{author}{\bibfnamefont{N.~N.} \bibnamefont{Ajitanand}},
  \bibinfo{author}{\bibfnamefont{J.~M.} \bibnamefont{Alexander}},
  \bibinfo{author}{\bibfnamefont{P.}~\bibnamefont{Chung}},
  \bibinfo{author}{\bibfnamefont{W.~G.} \bibnamefont{Holzmann}},
  \bibinfo{author}{\bibfnamefont{M.}~\bibnamefont{Issah}},
  \bibinfo{author}{\bibfnamefont{R.~A.} \bibnamefont{Lacey}},
  \bibinfo{author}{\bibfnamefont{A.}~\bibnamefont{Shevel}},
  \bibinfo{author}{\bibfnamefont{A.}~\bibnamefont{Taranenko}},
  \bibnamefont{and}
  \bibinfo{author}{\bibfnamefont{P.}~\bibnamefont{Danielewicz}},
  \bibinfo{journal}{Phys. Rev. C} \textbf{\bibinfo{volume}{72}},
  \bibinfo{pages}{011902(R)} (\bibinfo{year}{2005}),
  \urlprefix\url{https://link.aps.org/doi/10.1103/PhysRevC.72.011902}.

\bibitem[{\citenamefont{Sjöstrand et~al.}(2015)\citenamefont{Sjöstrand, Ask,
  Christiansen, Corke, Desai, Ilten, Mrenna, Prestel, Rasmussen, and
  Skands}}]{Sjostrand:2014zea}
\bibinfo{author}{\bibfnamefont{T.}~\bibnamefont{Sjöstrand}},
  \bibinfo{author}{\bibfnamefont{S.}~\bibnamefont{Ask}},
  \bibinfo{author}{\bibfnamefont{J.~R.} \bibnamefont{Christiansen}},
  \bibinfo{author}{\bibfnamefont{R.}~\bibnamefont{Corke}},
  \bibinfo{author}{\bibfnamefont{N.}~\bibnamefont{Desai}},
  \bibinfo{author}{\bibfnamefont{P.}~\bibnamefont{Ilten}},
  \bibinfo{author}{\bibfnamefont{S.}~\bibnamefont{Mrenna}},
  \bibinfo{author}{\bibfnamefont{S.}~\bibnamefont{Prestel}},
  \bibinfo{author}{\bibfnamefont{C.~O.} \bibnamefont{Rasmussen}},
  \bibnamefont{and} \bibinfo{author}{\bibfnamefont{P.~Z.}
  \bibnamefont{Skands}}, \bibinfo{journal}{Comput. Phys. Commun.}
  \textbf{\bibinfo{volume}{191}}, \bibinfo{pages}{159} (\bibinfo{year}{2015}),
  \eprint{1410.3012}.

\bibitem[{\citenamefont{Bellm et~al.}(2016)}]{Bellm:2015jjp}
\bibinfo{author}{\bibfnamefont{J.}~\bibnamefont{Bellm}} \bibnamefont{et~al.},
  \bibinfo{journal}{Eur. Phys. J.} \textbf{\bibinfo{volume}{C76}},
  \bibinfo{pages}{196} (\bibinfo{year}{2016}), \eprint{1512.01178}.

\bibitem[{\citenamefont{Reichelt et~al.}(2017)\citenamefont{Reichelt,
  Richardson, and Siodmok}}]{Reichelt:2017hts}
\bibinfo{author}{\bibfnamefont{D.}~\bibnamefont{Reichelt}},
  \bibinfo{author}{\bibfnamefont{P.}~\bibnamefont{Richardson}},
  \bibnamefont{and} \bibinfo{author}{\bibfnamefont{A.}~\bibnamefont{Siodmok}},
  \bibinfo{journal}{Eur. Phys. J.} \textbf{\bibinfo{volume}{C77}},
  \bibinfo{pages}{876} (\bibinfo{year}{2017}), \eprint{1708.01491}.

\bibitem[{\citenamefont{Gleisberg et~al.}(2009)\citenamefont{Gleisberg, Hoeche,
  Krauss, Schonherr, Schumann, Siegert, and Winter}}]{Gleisberg:2008ta}
\bibinfo{author}{\bibfnamefont{T.}~\bibnamefont{Gleisberg}},
  \bibinfo{author}{\bibfnamefont{S.}~\bibnamefont{Hoeche}},
  \bibinfo{author}{\bibfnamefont{F.}~\bibnamefont{Krauss}},
  \bibinfo{author}{\bibfnamefont{M.}~\bibnamefont{Schonherr}},
  \bibinfo{author}{\bibfnamefont{S.}~\bibnamefont{Schumann}},
  \bibinfo{author}{\bibfnamefont{F.}~\bibnamefont{Siegert}}, \bibnamefont{and}
  \bibinfo{author}{\bibfnamefont{J.}~\bibnamefont{Winter}},
  \bibinfo{journal}{JHEP} \textbf{\bibinfo{volume}{02}}, \bibinfo{pages}{007}
  (\bibinfo{year}{2009}), \eprint{0811.4622}.

\bibitem[{\citenamefont{Efron}(1979)}]{efron1979}
\bibinfo{author}{\bibfnamefont{B.}~\bibnamefont{Efron}}, \bibinfo{journal}{Ann.
  Statist.} \textbf{\bibinfo{volume}{7}}, \bibinfo{pages}{1}
  (\bibinfo{year}{1979}),
  \urlprefix\url{https://doi.org/10.1214/aos/1176344552}.

\bibitem[{\citenamefont{Khachatryan et~al.}(2016)}]{Khachatryan:2015lva}
\bibinfo{author}{\bibfnamefont{V.}~\bibnamefont{Khachatryan}}
  \bibnamefont{et~al.} (\bibinfo{collaboration}{CMS}), \bibinfo{journal}{Phys.
  Rev. Lett.} \textbf{\bibinfo{volume}{116}}, \bibinfo{pages}{172302}
  (\bibinfo{year}{2016}), \eprint{1510.03068}.

\bibitem[{\citenamefont{Aaboud et~al.}(2017)}]{Aaboud:2016yar}
\bibinfo{author}{\bibfnamefont{M.}~\bibnamefont{Aaboud}} \bibnamefont{et~al.}
  (\bibinfo{collaboration}{ATLAS}), \bibinfo{journal}{Phys. Rev.}
  \textbf{\bibinfo{volume}{C96}}, \bibinfo{pages}{024908}
  (\bibinfo{year}{2017}), \eprint{1609.06213}.

\bibitem[{\citenamefont{Khachatryan et~al.}(2017)}]{Khachatryan:2016txc}
\bibinfo{author}{\bibfnamefont{V.}~\bibnamefont{Khachatryan}}
  \bibnamefont{et~al.} (\bibinfo{collaboration}{CMS}), \bibinfo{journal}{Phys.
  Lett.} \textbf{\bibinfo{volume}{B765}}, \bibinfo{pages}{193}
  (\bibinfo{year}{2017}), \eprint{1606.06198}.

\bibitem[{\citenamefont{Voloshin and Zhang}(1996)}]{Voloshin:1994mz}
\bibinfo{author}{\bibfnamefont{S.}~\bibnamefont{Voloshin}} \bibnamefont{and}
  \bibinfo{author}{\bibfnamefont{Y.}~\bibnamefont{Zhang}}, \bibinfo{journal}{Z.
  Phys.} \textbf{\bibinfo{volume}{C70}}, \bibinfo{pages}{665}
  (\bibinfo{year}{1996}), \eprint{hep-ph/9407282}.

\bibitem[{\citenamefont{Poskanzer and Voloshin}(1998)}]{Poskanzer:1998yz}
\bibinfo{author}{\bibfnamefont{A.~M.} \bibnamefont{Poskanzer}}
  \bibnamefont{and} \bibinfo{author}{\bibfnamefont{S.~A.}
  \bibnamefont{Voloshin}}, \bibinfo{journal}{Phys. Rev.}
  \textbf{\bibinfo{volume}{C58}}, \bibinfo{pages}{1671} (\bibinfo{year}{1998}),
  \eprint{nucl-ex/9805001}.

\bibitem[{\citenamefont{Ackermann et~al.}(2001)}]{Ackermann:2000tr}
\bibinfo{author}{\bibfnamefont{K.~H.} \bibnamefont{Ackermann}}
  \bibnamefont{et~al.} (\bibinfo{collaboration}{STAR}), \bibinfo{journal}{Phys.
  Rev. Lett.} \textbf{\bibinfo{volume}{86}}, \bibinfo{pages}{402}
  (\bibinfo{year}{2001}), \eprint{nucl-ex/0009011}.

\bibitem[{\citenamefont{Adler et~al.}(2003)}]{Adler:2002tq}
\bibinfo{author}{\bibfnamefont{C.}~\bibnamefont{Adler}} \bibnamefont{et~al.}
  (\bibinfo{collaboration}{STAR}), \bibinfo{journal}{Phys. Rev. Lett.}
  \textbf{\bibinfo{volume}{90}}, \bibinfo{pages}{082302}
  (\bibinfo{year}{2003}), \eprint{nucl-ex/0210033}.

\bibitem[{\citenamefont{Adare et~al.}(2008)}]{Adare:2008ae}
\bibinfo{author}{\bibfnamefont{A.}~\bibnamefont{Adare}} \bibnamefont{et~al.}
  (\bibinfo{collaboration}{PHENIX}), \bibinfo{journal}{Phys. Rev.}
  \textbf{\bibinfo{volume}{C78}}, \bibinfo{pages}{014901}
  (\bibinfo{year}{2008}), \eprint{0801.4545}.

\bibitem[{\citenamefont{Aamodt et~al.}(2010)}]{Aamodt:2010pa}
\bibinfo{author}{\bibfnamefont{K.}~\bibnamefont{Aamodt}} \bibnamefont{et~al.}
  (\bibinfo{collaboration}{ALICE}), \bibinfo{journal}{Phys. Rev. Lett.}
  \textbf{\bibinfo{volume}{105}}, \bibinfo{pages}{252302}
  (\bibinfo{year}{2010}), \eprint{1011.3914}.

\bibitem[{\citenamefont{Sirunyan et~al.}(2018)}]{Sirunyan:2017pan}
\bibinfo{author}{\bibfnamefont{A.~M.} \bibnamefont{Sirunyan}}
  \bibnamefont{et~al.} (\bibinfo{collaboration}{CMS}), \bibinfo{journal}{Phys.
  Lett.} \textbf{\bibinfo{volume}{B776}}, \bibinfo{pages}{195}
  (\bibinfo{year}{2018}), \eprint{1702.00630}.

\end{thebibliography}

\end{document}